\documentclass[aps,prl,twocolumn,amsmath,amssymb]{revtex4-1}
\usepackage{graphicx}
\usepackage{tabularx}
\usepackage{color}

\newcommand{\Ang}{\mathrm{\AA}}
\newcommand{\ket}[1]{\bigl| {#1} \bigr\rangle}

\def\eV{\,\textrm{eV}}
\def\etal{\textit{et al.}}

\begin{document}

\title{Does Silicene on Ag(111) Have a Dirac Cone?}
\author{Yun-Peng Wang}
\author{Hai-Ping Cheng}
\affiliation{Quantum Theory Project and Department of Physics,
University of Florida, Gainesville, Florida 32611, USA}

\begin{abstract}
We investigate the currently debated issue of the existence of the
Dirac cone in silicene on an Ag(111) surface,
using first-principles calculations based on density functional theory
to obtain the band structure.
By unfolding the band structure in the Brillouin zone of a supercell to that of a primitive cell,
followed by projecting onto Ag and silicene subsystems,
we demonstrate that the Dirac cone in silicene on Ag(111) is destroyed.
Our results clearly indicate that the linear dispersions observed in both
angular-resolved photoemission spectroscopy (ARPES)
[P. Vogt \etal, Phys. Rev. Lett. 108, 155501 (2012)]
and scanning tunneling spectroscopy (STS)
[L. Chen \etal, Phys. Rev. Lett. 109, 056804 (2012)]
come from the Ag substrate and not from silicene.
\end{abstract}
\maketitle

Silicene, a two-dimensional (2D) allotrope of Si with a hexagonal
honeycomb-like lattice similar to graphene \cite{Nat.438.197},
has recently attracted intense attention.
It was proven in first-principles studies that low-buckled silicene is
thermally stable and has a linear electronic dispersion near
$K$ points at corners of the first Brillouin zone
\cite{PRB.50.14916, PRB.72.075420, CPL.379.81, PRL.102.236804},
which is similar to the behavior of graphene.
As a result, Si atoms in silicene are conjectured to be partially $sp^2$ hybridized
\cite{PRL.108.155501}.
Silicon nanostructures and silicene were successfully synthesized by depositing Si atoms on surfaces of Ag
\cite{PRL.108.155501,PRL.109.056804,NanoLett.8.271,APL.96.183102,APL.96.261905,APL.98.081909,APEX.5.045802,NANOLett.12.3507,JPCM.24.314211},
$\mathrm{ZrB_2}$ \cite{PRL.108.245501},
and recently Ir \cite{silicene-Ir}.
The Ag(111) substrate is ideal for growing silicene
because the tendency to form an Ag-Si alloy is low \cite{PRL.108.155501},
and as a result there are several different atomic arrangements for silicene on Ag(111) surfaces.
A linear dispersion relation was found in the $4\times 4$ structure
by Fleurence \etal{} \cite{PRL.108.245501}.
Feng \etal{} \cite{NANOLett.12.3507} reported three different phases
of silicene on Ag(111) surface, two with a $4\times4$
unit cell with respect to the Ag(111) lattice,
and a third phase, a $\sqrt{3}\times\sqrt{3}$ reconstruction
in reference to the low buckled silicene lattice, in which
they also found the existence of Dirac fermions \cite{PRL.109.056804}.
Atomic arrangements $\sqrt{13} \times \!\sqrt{13}\; R \,13.9^\circ$
of silicene on an Ag(111) surface were also observed \cite{APEX.5.045802}.
Various configurations were constructed and simulated numerically
using density functional theory (DFT) \cite{JPCM.24.314211}.

A common motivation of these studies was to detect and utilize the Dirac fermions in silicene.
Two groups have claimed to find evidence for the existence of Dirac fermions in silicene on Ag(111) surface.
In one experiment \cite{PRL.108.155501}, a linear dispersion near the Fermi energy in the silicene--Ag(111)
system was observed using angular-resolved photoemission spectroscopy (ARPES).
No such band was observed for a bare Ag(111) surface,
and so the linear dispersion was attributed to silicene.
The Dirac point was measured to be $ 0.3 \eV $ below the Fermi energy,
and the Fermi velocity was estimated to be
$1.3\times {10}^6 \, \mathrm{m \, s^{-1}}$ \cite{PRL.108.155501}.
In a second experiment \cite{PRL.109.056804},
the quasiparticle interference (QPI) patterns at the surface of the
silicene--Ag(111) system were observed using
scanning tunneling spectroscopy (STS).
The Dirac point was deduced from the linear dispersion curve to be
$\sim 0.5 \eV $ below the Fermi energy, and the Fermi velocity
is $\sim 1.2 \times 10^6\,\mathrm{m \, s^{-1}} $.
The Fermi velocities from these two experiments are close to the theoretical
prediction of $\sim 10^6\,\mathrm{m \,s^{-1}}$ \cite{PRL.102.236804}.
If judged merely from Fermi velocities, the experimentally observed
linear dispersions do coincide with the theoretical band structure.
However, the energy windows where the two experiments
\cite{PRL.108.155501,PRL.109.056804} observed a linear dispersion are too large.
Band structure from first-principle calculations \cite{PRL.102.236804}
showed linear dispersion only within a energy interval $ \pm 0.4\eV $,
which is several times smaller than the reported energy ranges
for linear dispersions, $ -3.0 \eV$ to $-0.3 \eV$ \cite{PRL.108.155501}
and  $ 0.4 \eV $ to $ 1.2 \eV$ \cite{PRL.109.056804}.
This discrepancy and the lack of dispersion measurements in the vicinity of and across the Dirac point make
the evidence for Dirac fermions in silicene reported by Ref.~\cite{PRL.109.056804} inconclusive.
A very recent paper \cite{PRL.110.076801} reported Landau level
measurements in silicene on Ag together with band structure calculations.
Absence of characteristic signals attributed to the Landau levels
disagrees with the experimental reports \cite{PRL.108.155501,PRL.109.056804}.
Band structure calculations showed that $p_z$ orbitals
are strongly hybridized with and delocalized into the Ag substrate.
The authors argued that the linear dispersion observed
in Ref.~\cite{PRL.108.155501} is not contributed by silicene,
and the only reasonable explanation is that
it comes from the Ag substrate \cite{PRL.110.076801};
however, there is no clear and straight-forward evidence presented.

In this Letter, we report results from DFT calculations that aim to understand the electronic structure of silicene on an Ag surface
in the context of the linear dispersion observed in experiments \cite{PRL.108.155501,PRL.109.056804}.
In all the above-mentioned work, DFT was routinely used to construct an atomic structural model
that reproduces the observed STM images
\cite{PRL.109.056804,APEX.5.045802,PRL.108.155501,PRL.108.245501,NANOLett.12.3507,JPCM.24.314211}.
Because of distortions of silicene on Ag surface, the unit cell of the silicene--Ag calculation is $3 \times 3$ times large
as primitive cell of silicene.
The resulting band foldings
make the calculated band structures in Ref.~\cite{PRL.110.076801} too complex to extract useful information.
Although Ref.~\cite{PRL.110.076801} argued that the experimentally observed
linear dispersions are not from silicene but from the Ag surface,
no direct connection was made between the experimental linear dispersions
and Ag bands.
In this work, we introduce a modified effective band structure (EBS)
technique to unfold bands from supercell calculations and thus make
tractable the identifying of signals from experiments with specific bands.
Unfolding the bands enables us to clearly identify the origins of the linear
dispersions reported in Refs.~\cite{PRL.108.155501} and \cite{PRL.109.056804}.
\begin{figure}[b]
\begin{center}
\includegraphics[width=\linewidth]{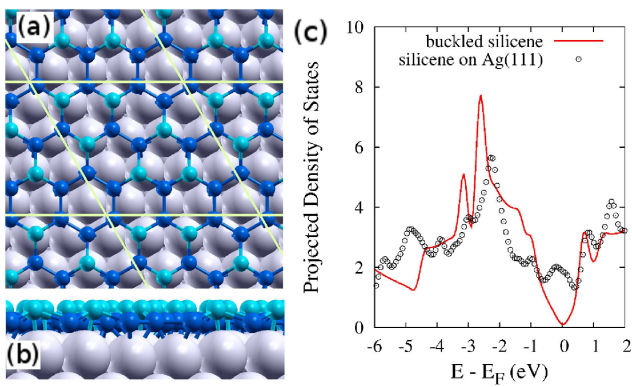}
\caption{
\label{fig:struct1}
(color online)
(a) Top and (b) side view of the silicene--Ag system (small balls: Si; big balls: Ag).
(c) (projected) density of states of low buckled silicene (line) and of silicene on Ag(111) surface (circles).
}
\end{center}
\end{figure}

We used the Vienna \textit{Ab initio} Simulation Package (VASP) \cite{PRB.59.1758},
a code based on a plane-wave basis and projector-augmented wave datasets \cite{PRB.59.1758}.
The Perdew-Burker-Ernzerhof form \cite{PRL.77.3865} of exchange-correlation functional was used in all the calculations.
The calculated bulk lattice constant of \textit{fcc} Ag is $ 4.174 \, \Ang $.
Although the atomic arrangements of silicene on Ag(111) surfaces are not unique
\cite{NANOLett.12.3507,APEX.5.045802,PRL.108.155501}, we focused on
the $4\times 4$ atomic arrangement since it is the most common structure \cite{JPCM.24.314211}.
The geometric configuration of silicene on an Ag(111) surface with a $4\times 4$ atomic arrangement
was constructed and optimized (see Fig.~\ref{fig:struct1}(a)(b)).
In one unit cell of the slab used for surface studies there are five Ag layers,
each with ($4\times 4$) Ag atoms.
Ag atoms in the bottom two layers were kept fixed at bulk lattice positions.
On the Ag(111) surface, the silicene sub-unit consists of 18 Si atoms.
After relaxation (with forces on atoms smaller than $ 0.01 \eV/\Ang $),
positions of six of the 18 Si atoms shift upward (away from Ag)
with respect to the other twelve, and the average distance between the
silicene sheet and the Ag surface is $ 2.44 \, \Ang $,
which is close to that reported in Ref.~\cite{PRL.108.155501}.
Geometric relaxation was also performed for standalone,
low-buckled silicene.

For optimized geometries we then calculated the density of states (DOS).
For standalone low-buckled silicene, the DOS at the Fermi energy is zero (see Fig.~\ref{fig:struct1}(c)).
Our symmetry analysis of wave functions showed that indeed at the Fermi energy
the $\pi$-bonding and anti-bonding bands do not overlap but touch each other at 
$K$ points at corners of the first Brillouin zone,
indicating that standalone silicene is a zero-gap semiconductor.
When silicene is placed on the Ag surface, the DOS projected onto the silicene
is no longer zero at the Fermi energy (see Fig.~\ref{fig:struct1}(c));
instead, a peak emerges near the Fermi energy as a result of silicene--Ag interaction.
The net charges on Ag and Si atoms are negligible (less than $0.03 \, e^-$
per atom according to Bader charge analysis \cite{bader}),
but the difference between the DOS of pristine silicene
and the DOS projected on silicene can not be describe by a simple rigid shift;
the interaction between Ag substrate and silicene is therefore
beyond charge doping.

Next, we looked into band structure, which offers more information on electronic structure than the DOS,
to compare the band structure of the silicene--Ag(111) system with that of standalone low-buckled silicene.
However, due to the distortion of silicene when deposited on Ag surfaces \cite{PRL.108.155501},
the unit cell is $3\times 3$ times the primitive unit cell of low-buckled
silicene \cite{PRL.102.236804}.
As a result of band foldings, a direct comparison between the band structure from supercell calculations
and that from primitive unit cell calculations is meaningless \cite{PRL.110.076801}.
Even if one performs a supercell calculation for the standalone silicene, it is not  trivial (nearly impossible) to extract information that can be compared with experimental data; band unfolding is necessary.

There are several options to unfold the band structure from supercell calculations.
First is to use the orbital-resolved spectral functions that can be calculated
using the Wannier function based unfolding method \cite{PRL.104.216401}.
This method is well suited to our system, but the Wannier functions
used in this method make it complicated to apply.
A second possibility that is free from the complications involving Wannier functions
is to use the effective band structure (EBS) method \cite{PRL.104.236403,PRB.85.085201}.
This method was originally used to study the effective dispersion in alloys;
however its original form is not ready to be used for our system because of its lack of atom and orbital resolutions.
We thus introduced a modified EBS method,
which enables calculations of orbital-projected spectral functions.

In the original EBS method \cite{PRB.85.085201}, the spectral function is defined as
\begin{equation}
\label{equ:tot_spectr}
\mathcal{A}(\vec{k}, E) = \sum_{N} P_{\vec{K}_jN}(\vec{k}) \, \delta(E_{\vec{K}_jN} - E) ,
\end{equation}
where the $P_{\vec{K}_jN}(\vec{k})$ are inner products of Kohn-Sham (KS)
wave functions $\ket{\vec{K}_j N}$ in the first Brillouin zone (BZ)
of the supercell and KS wave functions $\ket{\vec{k}_i n}$
in the first  BZ of the primitive cell;
$E_{\vec{K}_j N}$ are eigenenergies on predefined $k$-points $\{\vec{K}_j\}$
in the first BZ of the supercell system.
\begin{figure*}[t]
\begin{center}
\includegraphics[width=0.9\linewidth]{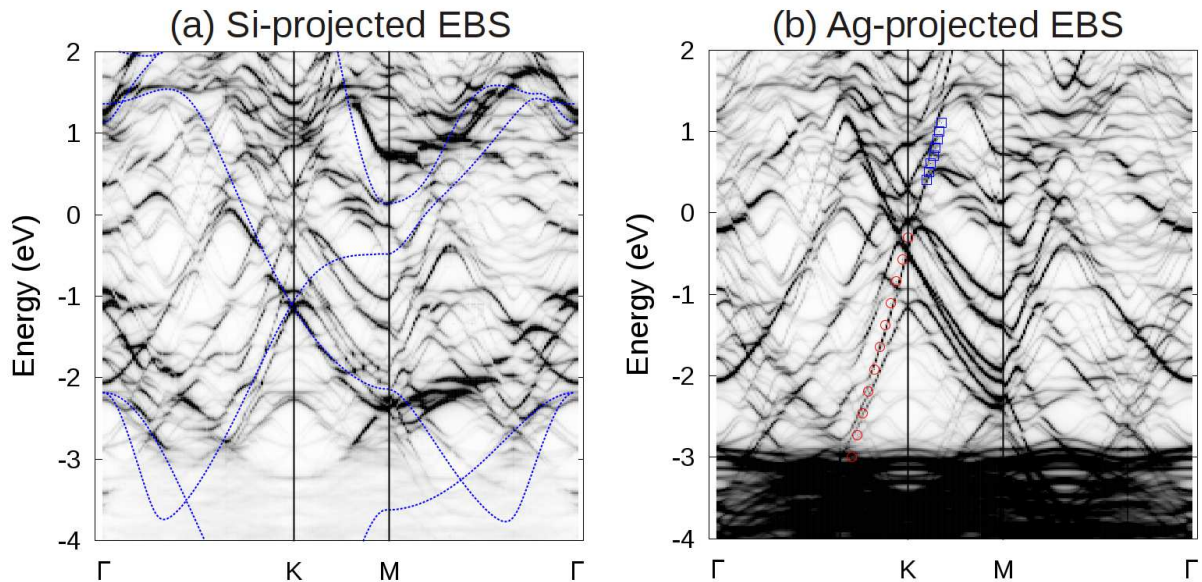}
\caption{
\label{fig:EBS}
 (Color online) Si- and Ag-projected effective band structure (EBS) of the silicene--Ag system.
(a) Si-projected effective band structures, with band structure of standalone low-buckled silicene
plotted as blue dashed lines.
Note that the band structure of low-buckled silicene was shifted downward by $ 1.1 \eV $.
(b) Ag-projected effective band structures with experimental observed linear dispersions
(red circles: measurements from Ref.~\cite{PRL.108.155501},
blue squares: measurements from Ref.~\cite{PRL.109.056804})}
\end{center}
\end{figure*}

To extend Eq.~(\ref{equ:tot_spectr}), we define the projected spectral functions
on specified atomic orbitals $\ket{\alpha}$ as
\begin{equation}
\mathcal{A}_{\alpha}(\vec{k}, E)
= \sum_{N} \bigl| p_{\alpha}^{\vec{K}_jN} \bigr|^2 \,
P_{\vec{K}_jN}(\vec{k}) \, \delta(E_{\vec{K}_jN} - E) ,
\end{equation}
where the $p_{\alpha}^{\vec{K}_jN}$ are projection functions defined as
\begin{equation}
p_{\alpha}^{\vec{K}_jN} = \bigl\langle {\vec{K}_jN} \big| {\alpha} \bigr\rangle
\end{equation}
i.e., expansion coefficients of KS wave functions $P_{\vec{K}_jN}(\vec{k})$ in atomic orbitals $\ket{\alpha}$.
With this extension, we calculated the spectral functions projected onto Ag and Si orbitals
with $\vec{k}$ along a high-symmetry path
$\Gamma$-$K$-$M$-$\Gamma$ in the first BZ
corresponding to the unit cell of standalone low-buckled silicene.
The calculated spectral functions were then summed into Ag- and Si-projected spectral functions
and divided by the number of atoms of a kind in one unit cell, so
their magnitudes are ready to compare.
These spectral functions are functions of $\vec{k}$ and $E$;
magnitudes of the spectral functions at $(\vec{k},E)$ are represented
in Fig.~\ref{fig:EBS} by different gray levels
(darker points represent larger spectral magnitude at that point).
Peaks in these spectral functions constitute continuous bands.

In Fig.~\ref{fig:EBS}(a), the band structure of standalone low-buckled silicene
is also plotted with the Si-projected spectral function as the blue dashed lines.
The band structure of standalone silicene was shifted $ 1.1 \eV $ downward
in order to match the Si-projected spectral functions  around $ -1.1 \eV $,
which suggests the interaction between Ag surface and silicene is more than charge doping.
The Ag $d$-bands dominate in the energy interval $-6 \eV $ through $-3\eV $,
so the Si-projected spectral function are hardly visible in this energy range (see Fig.~\ref{fig:EBS}(a)).
Most bands appear in both the Si- and Ag-projected spectral functions,
because the corresponding wave functions are neither entirely localized
in the Ag substrate nor in the silicene sheet.
Magnitudes of the Si- and Ag-projected spectral functions
can be compared in order to tell which atoms these bands come from.
The projected EBS shown in Fig.~\ref{fig:EBS} are somewhat complex,
because we use in our calculations a slab with finite thickness,
and the contributions from surface states mix with bulk states\cite{unfolding:slab}.

If one plots the band structure for a system with a Dirac cone along
high-symmetry path, the Dirac cone has two branches
touching each other at the $K$ point (Dirac point).
It seems from Fig.~{\ref{fig:EBS}(a)} that the Dirac point in silicene on the Ag substrate
moves to $-1.1 \eV $, because the upper branch of the Dirac cone in the $\Gamma$-$K$ portion
and the lower branch in the  $K$-$M$ portion coincide
with that of standalone low-buckled silicene moved downward by $ -1.1 \eV $.
However, compared to Fig.~{\ref{fig:EBS}(b)}, the lower branch of the Dirac cone
in the $\Gamma$-$K$ portion is not dominated by Si.
Most important, the upper branch of the Dirac cone in the K-M portion no longer exists.
These observations indicates that the Dirac cone in silicene on Ag has been destroyed.

\begin{figure}[t]
\begin{center}
\includegraphics[width=\linewidth]{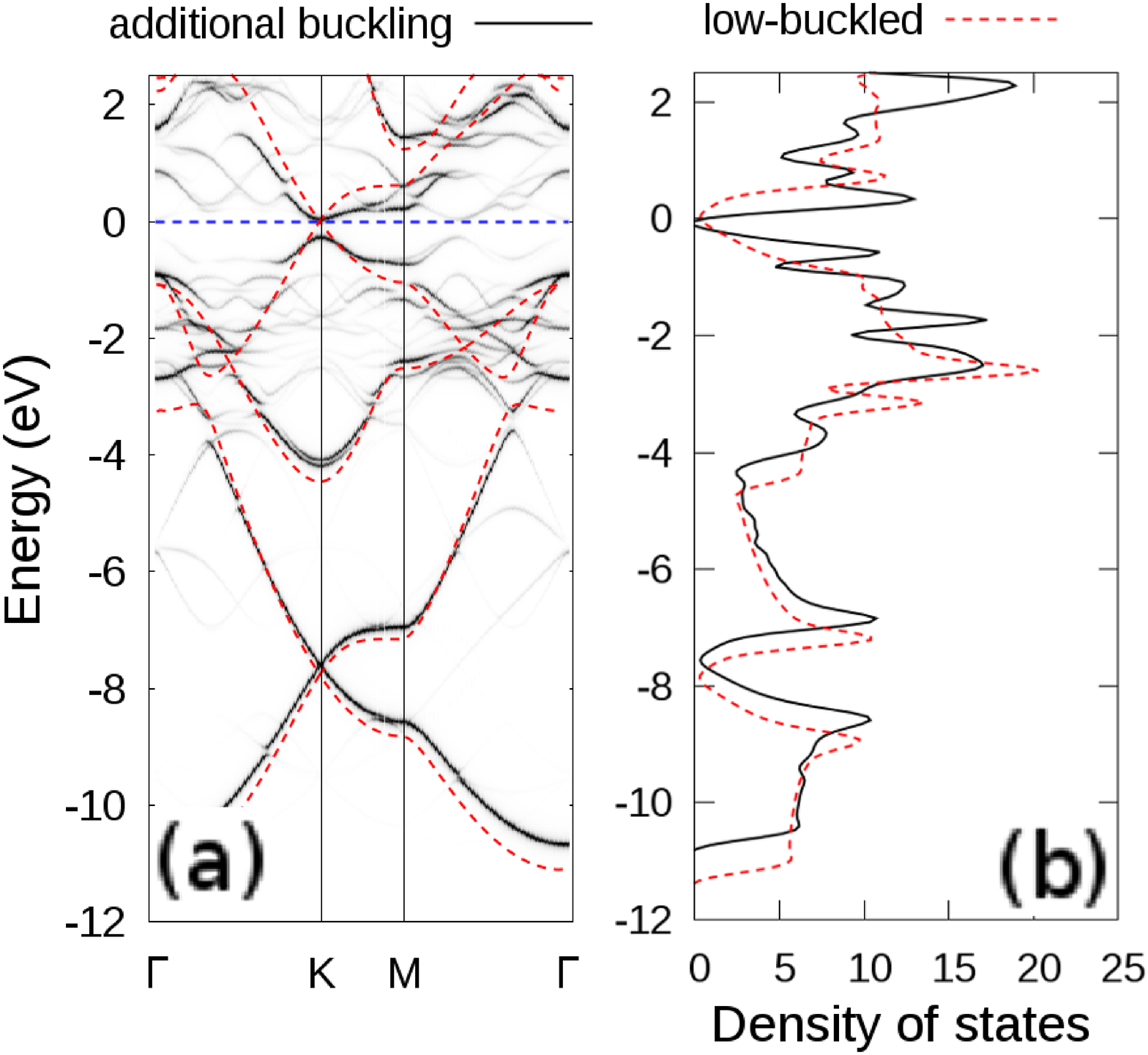}
\caption{
\label{fig:EBS-only-Si}
(Color online)
(a) Effective band structures and (b) density of states
of standalone silicene with additional buckling are compared
with that of low-buckled silicene (red dashed lines).
}
\end{center}
\end{figure}

The distortion of silicene on an Ag(111) surface and its interaction with Ag(111) surface
are two possible reasons for the disappearance of Dirac cone.
In order to show how the distortion affects the band structure
of silicene, we did another EBS calculation in which all the Ag atoms
in the unit cell were deleted, while silicon atoms were kept at their positions.
The effective band structure and density of states are shown
in Fig.~\ref{fig:EBS-only-Si}.
The most import observation from Fig.~\ref{fig:EBS-only-Si}
is the $\sim 0.3 \eV $ energy gap opening at the $K$ point.
The $\pi$-bonding and anti-bonding bands near the Fermi energy
do not touch each other any more, and develop to two flat bands in the
K-M path which correspond to the two peaks near the Fermi energy
in the DOS plot.
Our conclusion from this calculation is that the distortion seen
for silicene on Ag(111) surface alone can destroy the Dirac cone in silicene.

Finally, we discuss the comparison between our results and experimental observations.
Based on the (projected) EBS, we have concluded that the Dirac cone in silicene
is destroyed by the distortion of silicene and the interaction with Ag surface.
However, linear dispersions over very large energy ranges were observed
in experiments \cite{PRL.108.155501, PRL.109.056804}.
To understand this, we plotted the linear dispersions observed in
both experiments together with our Si-projected EBS of the silicene--Ag(111)
system, but we found no coincidence between the theory and the experiments.
We do find a connection when placing the experimental dispersion
functions on the Ag-projected EBS (see Fig.~\ref{fig:EBS}(b)):
we observe that the experimental dispersions coincide with
the $sp$-band of the Ag(111) surface.
Although $sp$-electrons in Ag are free-electron-like and the
$sp$-bands of Ag are actually parabolic,
fitting to a linear function is acceptable for a finite energy interval
away from its minimum energy.
The large energy range where the $sp$-band of Ag appears
is consistent with that in which the experiments observed linear dispersions,
and provides the only plausible explanation.
Additionally, below  $-3\eV $, the $sp$-bands of Ag are buried by $d$-bands,
which can explain why the lower bound of the linear dispersion observed
in Ref.~\cite{PRL.108.155501} is  $ -3 \eV $.
Overall, based on our calculations we reach a different understanding
of the experimental observed linear dispersions:
they are from Ag(111) surface instead of from silicene.

In summary, inspired by experimental observations of linear dispersions
in silicene--Ag(111), we conducted band structure calculations using DFT
and calculated the projected effective band structures.
The Dirac cone in silicene was observed to be destroyed
by distortions and interactions with Ag surface.
The linear dispersions observed in experiments were found to coincide
with the Ag-projected but not Si-projected EBS,
which indicates that these linear dispersions come from the Ag surface
instead of from silicene.
We presented a different interpretation of the experimental linear
dispersions, attributed to be from the Ag surface instead of from silicene.

This work was supported by the US Department of Energy
(DOE), Office of Basic Energy Sciences (BES), under Contract
No. DE-FG02-02ER45995.
The computation was done using the utilities of the
National Energy Research Scientific Computing Center (NERSC).
Authors also would like to thank T. Berlijn for fruitful discussion.


\end{document}